\documentstyle[aps,preprint,pra,epsfig]{revtex}

\newcommand{\be}{\begin{equation}}
\newcommand{\en}{\end{equation}}
\newcommand{\bea}{\begin{eqnarray}}
\newcommand{\ena}{\end{eqnarray} }

\newcommand{\ii}{{\bf i}}

\tightenlines

\begin{document}
\title{Phase diagram of Josephson junction arrays with capacitive disorder}
\author{F. P. Mancini, P. Sodano, and A. Trombettoni}
\address{Dipartimento di Fisica and Sezione I.N.F.N., Universit\`a di \\
Perugia, Via A. Pascoli, I-06123 Perugia, Italy}
\date{\today }
\maketitle

\begin{abstract}
We study the phase diagram at finite temperature of Josephson junction
arrays with capacitive disorder (i.e., random offset charges and/or random
charging energies): in the limit of large particle numbers per junction,
this is a remarkable physical realization of the disordered boson Hubbard
model. By using a mean-field approximation, 
we compute the average free energy and the equation for the
phase boundary line between the insulating and the superconducting phase. We
find that the Mott-insulating lobe structure disappears for large variance ($%
\sigma \gtrsim e$) of the offset charges probability distribution. Further,
with nearest-neighbor interactions, the insulating lobe around $q=e$ is
destroyed even for small values of $\sigma$. In the case of random charging
energies, until the variance of the distribution reaches some critical value
the superconducting phase increases in comparison to the situation in which
all self-capacitances are equal.
\end{abstract}




\section{INTRODUCTION}

In practical realizations of Josephson devices \cite{fazio01}, one has to
deal with capacitance disorder caused either by offset charge defects in the
junctions or in the substrate (random offset charges) \cite{krupenin00} or
by imperfections in the construction of the devices, which may lead to
random capacitances of the Josephson junctions. Random offset charges cannot
be made to vanish by using a gate for each superconducting island since in
large arrays too many electrodes would be necessary (i.e., too complicated
fabrication procedures). In principle, an external uniform charge can be
introduced and tuned in Josephson junctions arrays (JJA's) by applying a gate
voltage with respect to the ground plane: in Ref.\cite{lafarge95} this
situation was analyzed experimentally by placing a gate underneath a
Josephson array and it was observed a sensible variation ($\sim 40\%$) of
the resistance between the unfrustrated and the fully frustrated array.
Although from a theoretical point of view charge and magnetic frustrations
are dual to each other, experimentally it is possible to tune only the
magnetic frustration in a controlled way. 
For this reason, it is widely believed that
a challenging task for the theory is to develop reliable techniques to
investigate the effects of random charge frustration on the phase structure
of the arrays.

In this paper, we shall address the problem of determining the finite
temperature phase diagram of JJA's with capacitive disorder (i.e., with random
offset charges and/or random self-capacitances). To derive the phase
boundary between the insulating and the superconducting phase, we shall use
a mean-field (MF) theory approach 
in the path-integral approach for quantum JJA's
with offset charges and general capacitance matrices 
\cite{kopec00,grignani00}. We find that the charge disorder supports
superconductivity and that the relative variations of 
the insulating and superconducting regions
depend on the mean value $q$ of the charge probability distribution: when 
$q=0$, increasing the disorder leads to an enlargement of 
the superconducting phase. 
If the charge disorder is sufficiently strong ($\sigma \gtrsim e$),
the lobe structure \cite{fazio01} disappears: in other words, the phase
boundary line (and the correlation functions) does
not depend any longer on $q$. 
In the following, we shall provide a quantitative analysis of this
phenomenon. We shall consider Gaussian, uniform (as in Ref.\cite{fisher89})
and $\delta$-like distributions. For $T=0$
our results agree with the results obtained in 
Refs.\cite{fisher89} and \cite{vanotterlo93}.
Also the randomness of the self-capacitances leads
to remarkable effects: the superconducting phase increases with respect to
the case where disorder is not present. We shall also compare the results 
of the functional MF approach 
with the molecular MF discussed in Appendix B.

The Hamiltonian commonly used to describe the Cooper pair tunneling in
superconducting quantum networks defines the so called quantum phase model
(QPM). In its most general form it is given by 
\begin{equation}
H=\frac{1}{2}\sum_{ij}(Q_{{\bf i}}+q_{{\bf i}})C_{{\bf i}{\bf j}}^{-1}(Q_{%
{\bf j}}+q_{{\bf j}}) -E_{J}\sum_{\left \langle ij\right\rangle }\cos
(\varphi_{{\bf i}}-\varphi_{{\bf j}})  \label{QPM}
\end{equation}
where $\varphi_{{\bf i}}$ is the phase of the superconducting order
parameter at grain ${\bf i}$. Its conjugate variable ${n}_{{\bf i}}$ ($[{
\varphi_{{\bf i}}}, {n}_{{\bf i}}]=i\:\delta_{{\bf i} {\bf j}}$) describes
the number of Cooper pairs in the ${\bf i}$th superconducting grain. The
symbol $\left\langle ij\right\rangle$ indicates a sum over nearest-neighbor
grains only. The second term in the Hamiltonian (\ref{QPM}) describes the
hopping of Cooper pairs between neighboring sites ($E_J$ is the Josephson
energy). The first term determines the electrostatic coupling between the
Cooper pairs: ${Q}_{{\bf i}}$ is the excess of charge due to Cooper pairs ($%
Q_{{\bf i}}=2e n_{{\bf i}}$) on site ${\bf i}$ and $C_{{\bf i} {\bf j}}$
is the capacitance matrix. An external gate voltage $V_{{\bf i}}$ gives the
contribution to the energy via the offset charge $q_{{\bf i}}=\sum_{{\bf j}}
C_{{\bf i}{\bf j}} V_{{\bf j}}$. This external voltage can be either applied
to the ground plane or, more interestingly, it may be induced by charges
trapped in the substrate. In the latter situation $q_{{\bf i}}$ is a random
variable: the effects of this randomness on the phase diagram are the main
object of our investigation. We shall also treat explicitly the case of
random self-capacitance $C_{{\bf i}{\bf i}}$ which correspond to a random
charging energy $E_C=e^2 C_{{\bf i}{\bf i}}^{-1}/2$.

As is well known \cite{fazio01}, the QPM (\ref{QPM}) is equivalent to the
boson Hubbard model (BHM) in the limit of large particle numbers per
junction. The BHM describes soft-core bosons hopping on a lattice \cite%
{fisher89} and is defined as 
\begin{equation}
H=\frac{1}{2}\sum_{ij} n_{{\bf i}} U_{ij} n_{{\bf j}}- \mu \sum_{i} n_{{\bf i
}} - \frac{t}{2} \sum_{\left \langle ij\right\rangle } (b_{{\bf i}}^{\dag}b_{
{\bf j}}+H.C.).  \label{BHM}
\end{equation}
Here, $b_{{\bf i}}^{\dag}$ ($b_{{\bf i}}$) is the creation (annihilation)
operator for bosons and $n_{{\bf i}}=b_{{\bf i}}^{\dag}b_{{\bf i}}$ is the
number operator. By writing the field $b_{{\bf i}}$ in terms of its
amplitude and phase and by neglecting the deviations of the amplitude from
its average, we are lead to the QPM (\ref{QPM}). An exact mapping between
the two models has been derived in Ref. \cite{anglin01}. The hopping term is
associated with the Josephson tunneling ($\langle n\rangle t \to E_J$)
whereas $U_{{\bf i}{\bf j}} \to 4e^2 C_{{\bf i}{\bf j}}^{-1}$ describes the
Coulomb interactions between bosons. The chemical potential in the BHM plays
a role analogous to the one of the external charge in the QPM ($\mu \to q_{%
{\bf i}}$). Thus a QPM with random offset charge corresponds to a BHM with
random on-site energies.

The disordered BHM has attracted much attention during the last decade \cite%
{fisher89,pazmandi95,kisker97,pazmandi97,herbut98,granato98,lee01}. In the
pioneering work of Fisher {\em et al.} \cite{fisher89} the phase diagram at $%
T=0$ was studied. Without disorder, the BHM exhibits two types of phases: a
superfluid phase and a Mott insulating phase, with the latter characterized
by integer (or commensurate) boson densities, by the existence of a gap for
the particle-hole excitations, and by zero compressibility. The phase
structure is determined by the two competing terms of the Hamiltonian: the
charging energy leads to a charge localization in the array, while the
Josephson energy induces a phase coherence giving rise to the overall array
superfluidity. When disorder is present, 
a third, intermediate, phase occurs: the Bose
glass (BG). This phase has an infinite superfluid susceptibility, but no gap
and finite compressibility. It was early realized that considering
uniform probability distributions there is no BG in the MF \cite%
{fisher89}. 
However, it has been shown that
introducing a not uniform triangular probability distribution of offset
charges the BG phase appears in the MF 
phase diagram \cite{pazmandi95}. 
A direct Mott insulator to superfluid transition without an intervening BG
phase for weak disorder has been recently investigated in two dimensions 
\cite{kisker97,pazmandi97}. Very recently, the phase diagram of the
two-dimensional disordered BHM has been studied at $T=0$ by means of Monte
Carlo simulations \cite{lee01} evidencing the existence of a Bose glass to
superfluid transition in the strong-disorder regime.

In the present paper we shall consider uniform 
disorder distributions: our MF
approach distinguishes the phases with order parameter $\left\langle \cos{%
\varphi_{{\bf i}}}\right\rangle$ equal to zero (insulating region) or 
different from zero (superconducting region), but it cannot capture the BG. 
Therefore, the full phase diagram
is much richer than the MF one. 
It is obvious that for $d=1$ (e.g. JJ chains \cite{bobbert92}) 
the MF fails to provide reliable information.
However, the MF allows us to take a first step towards
the understanding of
the properties of the disordered JJA and BHM at finite temperature.
Furthermore, conventional wisdom suggests that for large dimensions
the MF approach provides the correct phase 
diagram of the system.

The plan of the paper is the following: in Section II, we outline the
MF theory for the pure quantum JJA and then we compute the free
energy averaged over the disorder. In this way we are able to get a general
formula for the phase boundary line at finite temperature. 
Section III is devoted to the study of the effects of random offset 
charges with diagonal and
nearest-neighbor capacitance matrices. In Section IV 
we consider random self-capacitances. 
To check our results, we study in Appendix A the infinite-range hopping
limit and provide in Appendix B an alternative and more intuitive
MF approach for JJA's in the presence of capacitive
disorder. Finally, Sec. V is
devoted to concluding remarks.

\section{AVERAGE OVER THE DISORDER IN MEAN-FIELD THEORY}

In a functional approach which makes use of the Hubbard-Stratonovich
transformation, the partition function of quantum JJA's may be written as
\cite{grignani00} \begin{equation} Z = \int \prod_{{\bf i}} D\psi_{{\bf i}}
D\psi_{{\bf i}}^* e^{ \int_{0}^{\beta} d\tau (-\frac{2}{E_J} \sum_{{\bf i}
{\bf j}} \psi_{{\bf i} }^* \gamma_{{\bf i} {\bf j}}^{-1} \psi_{{\bf j}} )}
e^{-S_{Eff}[\psi]} \label{13}
\end{equation}
where $\beta=1/k_B T$ and 
$\gamma_{{\bf i} {\bf j}}=1$ if ${\bf i},{\bf j}$ are nearest
neighbors and equals zero otherwise (i.e., the hopping term just between
nearest neighbors). If $\gamma_{{\bf i} {\bf j}}=1$ for all pairs ${\bf
i},{\bf j}$ on the lattice, we are led to the infinite-range hopping limit
which provides a remarkable example of exactly solvable MF theory \cite%
{fisher89}. In the following we shall treat explicitly also the
infinite-range case. In Eq. (\ref{13}), $S_{eff}$ is the effective action
for the auxiliary Hubbard-Stratonovich field $\psi_{{\bf i}}$:
\begin{equation}
\label{14}
S_{Eff}\left[\psi \right]= - \ln \left\{ \int \prod_{{\bf i}} D
\varphi_{{\bf i}} \: \: \exp \left[ \int_{0}^{\beta} d\tau \left(
-\frac{1}{2}\sum_{{\bf i } {\bf j}} C_{{\bf i} {\bf
j}}\frac{\dot{\varphi_{{\bf i}}}}{2e} \frac{\dot{ \varphi_{{\bf j}}}}{2e} +
i \sum_{{\bf i}} ( q_{{\bf i}} \frac{\dot{\varphi_{ {\bf
i}}}}{2e}-\psi_{{\bf i}} e^{i\varphi_{{\bf i}}}-\psi_{{\bf i}}^* e^{-i
\varphi_{{\bf i}}})\right) \right]\right\} .
\end{equation}

The field $\psi_{{\bf i}}$ may be regarded as the order parameter for the
insulator-superconductor phase transition because it turns out to be
proportional to $\langle e^{i\varphi_{{\bf i}}}\rangle$. Since the phase
transition is second-order \cite{simkin96}, close to the onset of
superconductivity the order parameter $\psi_{{\bf i}}$ is small. One may
then expand the effective action up to the second order in $\psi_{{\bf i}}$,
getting 
\begin{equation}  \label{part-funct}
Z =\int \prod_{{\bf i}} D\psi_{{\bf i}} D\psi_{{\bf i}}^* e^{-F[\psi]};
\end{equation}
$F[\psi]$ is the Ginzburg-Landau free energy, which - to the second order in 
$\psi_\ii$ - is given by 
\begin{equation}  \label{prima_f}
F[\psi]=\int_{0}^{\beta} d\tau \int_{0}^{\beta} d\tau ^{\prime}\sum_{{\bf i} 
{\bf j}} \psi_{{\bf i}}^*(\tau) \left[ \frac{2}{E_J}\gamma_{{\bf i} {\bf j}%
}^{-1} \delta(\tau-\tau^{\prime})- G_{{\bf i} {\bf j}}(\tau,\tau^{\prime})%
\right]\psi_{{\bf j}}(\tau^{\prime})
\end{equation}
where the phase correlator $G_{{\bf i} {\bf j}}$ is given by
\begin{equation}
 \label{16}
G_{{\bf i} {\bf j}}(\tau, \tau^{\prime})=\frac{1}{\beta^2} \:
\frac{ \delta^2S_{Eff}[\psi]}{
\delta \psi_{{\bf i}}(\tau) \delta \psi_{{\bf j}} (\tau^{\prime})} \Bigg|
_{\psi,\psi^* = 0} =\langle e^{i\varphi_{{\bf i}}(\tau) -i \varphi_{{\bf j
}(\tau^{\prime})}} \rangle_0.
\end{equation}
A straightforward, but lengthy, calculation gives \cite{grignani00} 
\[
G_{{\bf r}{\bf s}}(\tau;\tau^{\prime})=\delta_{{\bf r} {\bf s}}e^{
-2e^2C^{-1}_{{\bf r} {\bf r}} |\tau-\tau^{\prime}|} \>\> \cdot \> 
\]
\begin{equation}  \label{phase-corr}
\cdot \> \frac{\sum_{[n_{{\bf i}}]}e^{ -\sum_{{\bf i} {\bf j}} 2e^2\beta C_{
{\bf i} {\bf j}}^{-1}(n_{{\bf i}}+\frac{q_{{\bf i}}}{2e})(n_{{\bf j}}+ \frac{
q_{{\bf j}}}{2e})-\sum_{{\bf i}}4e^2C_{{\bf r} {\bf i}}^{-1}(n_{{\bf i}}+ 
\frac{q_{{\bf i}}}{2e})(\tau-\tau^{\prime})} }{\sum_{[n_{{\bf i}}]}e^{-\sum_{
{\bf i} {\bf j}} 2 \beta e^2C_{{\bf i} {\bf j}}^{-1}(n_{{\bf i}}+\frac{q_{
{\bf i}}}{2e})(n_{{\bf j}}+ \frac{q_{{\bf j}}}{2e}) } }.
\end{equation}
Here, $n_{{\bf i}}$ assumes all integer values and $\sum_{[n_{{\bf i}}]}$ is
a sum over all the configurations. If one introduces the Fourier transforms
in the imaginary time and in the space 
\begin{eqnarray}
\psi_{{\bf i}}(\tau)&=&\frac{1}{\beta N}\sum_{{\bf k} \mu} \psi_{{\bf k}
}(\omega_{\mu})e^{i({\bf k}\cdot {\bf i}+\omega_\mu \tau)},
 \label{28}
 \\
G_{{\bf i}}(\tau)&=&\frac{1}{\beta N}\sum_{{\bf k} \mu} G_{{\bf k}
}(\omega_{\mu})e^{i({\bf k}\cdot {\bf i}+\omega_\mu \tau)}
 \label{giq}
\end{eqnarray}
(with $\omega_{\mu}$ Bose-Matsubara frequencies and ${\bf k}$ vectors of the
reciprocal lattice), the Ginzburg-Landau free energy (\ref{prima_f}), reads 
\cite{vanotterlo93,panyukov89}
\begin{equation}
F[\psi]= \frac{1}{\beta N}\sum_{\mu {\bf k}{\bf k}^{\prime}}\psi^*_{{\bf k}%
}(\omega_{\mu}) \left[\frac{2}{E_J} \gamma_{{\bf k}}^{-1}\delta_{{\bf k}{\bf %
k}^{\prime}}-\frac{G_{{\bf k}-{\bf k}^{\prime}} (\omega_{\mu})}{N}\right]
\psi_{{\bf k}^{\prime}}(\omega_{\mu}),
\label{seconda_f}
\end{equation}
where 
\begin{equation}  \label{gamma_fourier}
\gamma_{{\bf i} {\bf j}}^{-1}=\frac{1}{N}\sum_{{\bf k}} \gamma_{{\bf k}%
}^{-1}e^{i{\bf k}\cdot( {\bf i}-{\bf j})}.
\end{equation}
$\gamma_{{\bf k}}^{-1}$ is the inverse of the Fourier transform of the
Josephson coupling strength $\gamma_{{\bf i}{\bf j}}$; since $\gamma_{{\bf i}
{\bf j}}=1$ if ${\bf i},{\bf j}$ are nearest neighbors and zero otherwise,
we have $\gamma_{{\bf k}}^{-1}=(\sum_{{\bf p}}e^{-i {\bf k}\cdot {\bf p}
})^{-1}$, where ${\bf p}$ is a vector connecting two nearest-neighbors
sites. Expanding in ${\bf k}$ one gets $\gamma_{{\bf k}}^{-1}=1/z + \cdots$,
where $z$ is the coordination number. Substituting in Eq. (\ref{seconda_f})
and keeping only the lowest-order terms in $\omega_\mu$, ${\bf k}$ and
$1/z$, the MF Ginzburg-Landau free energy reads \begin{equation}
\label{ginzburg-landau-mf} F[\psi]\simeq \frac{1}{\beta N}\sum_{{\bf k} \mu}
\bigg[ \frac{2}{E_J z}-G_{{\bf 0}} +\cdots \bigg] |\psi_{{\bf
k}}(\omega_{\mu})|^2. \end{equation}
In Eq. (\ref{ginzburg-landau-mf}), $G_{{\bf 0}}$ is
\begin{equation}  \label{Gcon0}
G_{{\bf 0}}=\frac{1}{N}\sum_{{\bf r}}G_{{\bf r}}(\omega_{\mu}=0,T=T_c).
\end{equation}
As evidenced in Ref.\cite{grignani00} the MF theory approximation
amounts to neglect all higher order terms in Eq. (\ref{ginzburg-landau-mf}).

For a given realization of the disorder, 
the Ginzburg-Landau free energy (i.e., the free energy near the transition) 
is given by Eq. (\ref{ginzburg-landau-mf}). The average of the free 
energy over all the possible realization of the disorder allows for
evaluation of the effect of a random charge frustration $\{q_{{\bf i}}\}$ or
a random diagonal charging energy $U_{{\bf i} {\bf i}}=4e^2 C_{{\bf i}{\bf
i}}^{-1}$: one has then \begin{equation}  \label{av-ginzburg-landau}
\bar{F}[\psi]=\int d\{X\} P(\{X\}) F[\psi],
\end{equation}
where $P(\{X\})$ is a given probability distribution and $d\{X\}
P(\{X\})=\prod_{{\bf i}} dq_{{\bf i}} P(q_{{\bf i}})$ when one considers
random offset charges or $d\{X\} P(\{X\})=\prod_{{\bf i}} dU_{{\bf i}{\bf i}%
} P(U_{{\bf i}{\bf i}})$ for random charging energies. The random variables
on different sites are taken to be independent. The phase boundary line
between the insulating and the superconducting phase is determined by
requiring that $\bar{F}=0$, which in turn leads to 
\begin{equation}  \label{boundary-line}
1=z\frac{E_J}{2} \bar{G_{{\bf 0}}}.
\end{equation}
In Appendix A we will discuss how Eq. (\ref{boundary-line})
is modified in the infinite-range hopping limit. 
A comparison of the MF discussed in this Section with 
an alternative MF approach in the presence of disorder is given
in Appendix B.

\section{RANDOM OFFSET CHARGES}

In the following, we shall consider three different random offset charges
probability distribution with mean $q$ and width $\sigma$. That is, a
Gaussian distribution $P(q_{{\bf i}})=$ const $\times \cdot e^{-(q_{{\bf i}
}-q)^2/2\sigma^2}$, a uniform distribution $P(q_{{\bf i}})=$ const between
$ q-\sigma$, and $q+\sigma$ and a sum of $\delta$-like distributions
$P(q_{{\bf i}})=\sum_n p_n \delta(q_{{\bf i}}-ne)$, with $\sum_n p_n=1$. For
a diagonal capacitance matrix, Eq. (\ref{boundary-line}) leads to
\begin{equation}
\frac{1}{\alpha }=\int dq P(q)g(q,y) ,
\label{boundary-line-charge}
\end{equation}
with $\alpha =zE_{J}/4E_{c}$, $y=k_{B}T_{c}/E_{c}$, and
\begin{equation}
g(q,y)=\frac{\sum_{n}e^{-\frac{4}{y}(n+q/2e)^{2}}\frac{1}{1-4(n+q/2e)^{2}}}{%
\sum_{m}e^{-\frac{4}{y}(m+q/2e)^{2}}}.  \label{g_qy}
\end{equation}
We observe that the lobes of JJA are invariant under $q/2e \to q/2e +1$ and
symmetric around $q/2e=n+1/2$, where $n$ is an integer: 
\begin{equation}
g(q+2ne,y)=g(q,y); \, \, g(n+1/2+q/2e,y)=g(n+1/2-q/2e,y).  \label{per}
\end{equation}
If one considers the infinite-range hopping limit, one still gets Eq. (\ref%
{boundary-line-charge}) provided that $\alpha=J/4E_{c}$.

\begin{figure}[h]
\centerline{\psfig{figure=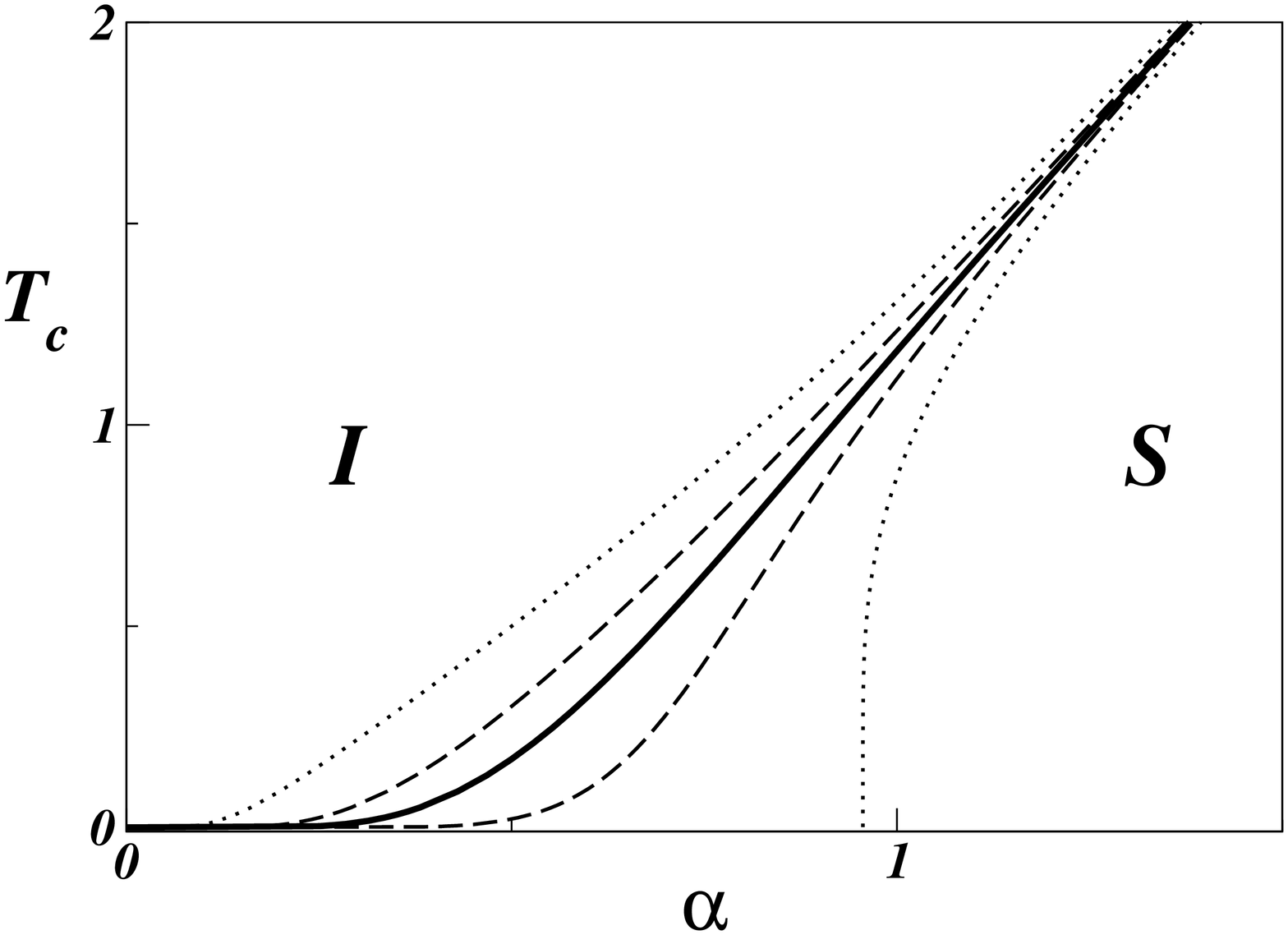,width=72mm,angle=270}}
\caption{Phase diagram for random offset charges with Gaussian
distribution ($T_c$ is in units of $k_B /E_C$). The bold line is for
$\sigma=e$ and it represents the case of large variance 
and it is found for all values of the mean $q$. To 
the left (right), we plot $q=e$ ($q=0$); we use $\sigma/2e=0.1$ 
(dotted line) and $0.25$ (dashed line). The {\bf  I} and {\bf  S}
indicate, respectively, insulating and superconducting phase.}
\label{figure1}
\end{figure}

The results obtained from Eq. (\ref{boundary-line-charge}) with a Gaussian
distribution are displayed in Fig. \ref{figure1}. One observes that when $%
q/2e=0 $, increasing $\sigma $ favors the superconducting phase while, when $%
q/2e=1/2$, increasing $\sigma $ leads to the increase of the insulating
phase. For large $\sigma$ (i.e., $\sigma \gtrsim e$), the phase boundary
line is the same for all the values of $q$ (in Fig. \ref{figure1} the
large-$\sigma$ behavior is represented by the bold line). This is expected
since, when $ \sigma$ is large, the average free energy $\bar{F}$ does not
depend any longer on $q$.

\begin{figure}[h]
\centerline{\psfig{figure=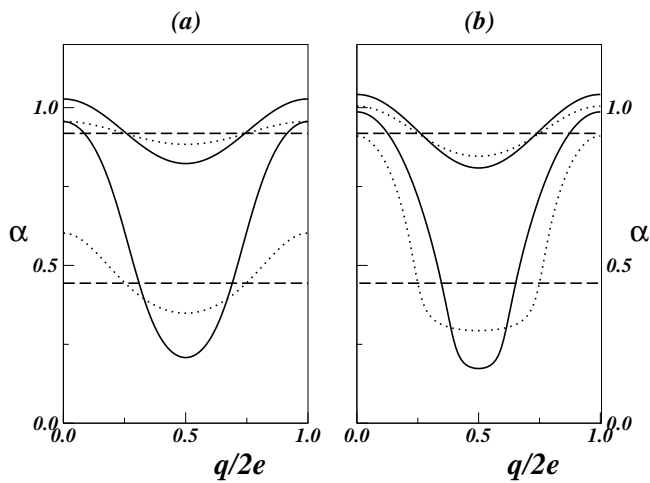,width=72mm,angle=270}}
\caption{Phase diagram with diagonal capacitance and 
random offset charges with Gaussian (a) and uniform (b)
distribution. Top (bottom) of the figures: $k_B T /E_C=1 (0.1)$. We 
plot the cases $\sigma/2e=0.1$ (solid lines), $0.25$ (dotted lines), $0.5$
(dashed lines). For large $\sigma/2e$ the phase boundary line is flat and
it is the same for both distributions.}
\label{figure2}
\end{figure}

\begin{figure}[h]
\centerline{\psfig{figure=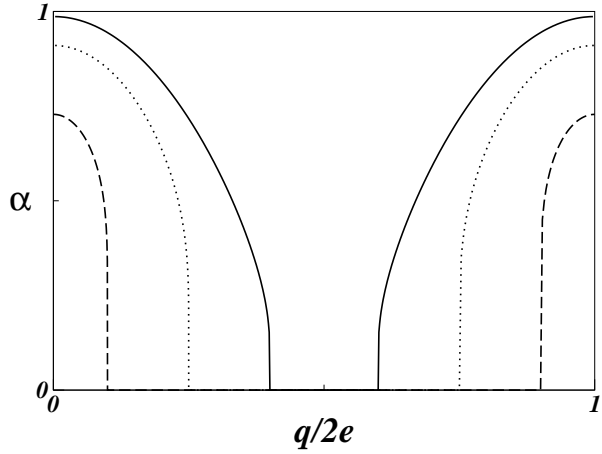,width=72mm,angle=270}}
\caption{Phase diagram at $T=0$ for random offset charges with
uniform distribution and short-ranged inverse capacitance matrix.
$\sigma/2e$ is respectively
$0.1$ (solid line), $0.25$ (dotted line) and $0.40$ (dashed line).}
\label{figure3}
\end{figure}

A useful representation of the phase diagram is obtained by plotting, at
fixed temperature, the phase boundary line on the plane $q$-$\alpha$.
Without disorder one observes the well-known lobe structure \cite{fazio01}.
In the presence of weak disorder and at $T=0$, the lobes shrink, evidencing
a decrease of the insulating phase: for a Gaussian (or unbounded)
distribution the insulating phase completely disappears even for an
arbitrarily weak disorder \cite{fisher89}. In Fig. \ref{figure2} we exhibit
the phase boundary line on the plane $q-\alpha$ at finite temperature for
the Gaussian and uniform distributions. Of course, even at finite
temperature, when the disorder increases the lobes flatten and the same lobe
structure is obtained from both distributions. At $T=0$ we recover the
result of Ref. \cite{fisher89}. This can be easily seen if one observes
that, at low temperatures, for $|q| < e$, one has \cite{vanotterlo93}
\begin{equation}
g(q,y \to 0) = \frac{1}{1-4(\frac{q}{2e})^2}.
\end{equation}
Without disorder ($\sigma=0$), Eq. (\ref{boundary-line-charge}) simply gives 
$\alpha=1-4(q/2e)^2$. With the Gaussian distribution, since $g$ has a pole
in the half-integer value of the Cooper charge, the integral in Eq. (\ref%
{boundary-line-charge}) diverges, and $\alpha \to 0$; i.e., the lobes
disappear for every value of $\sigma$. As evidenced in Fig. \ref{figure3}, for
a uniform distribution, when $\sigma > e$, then $\alpha \to 0$; when $%
\sigma < e$, $\alpha\to 0$ only for $e-\sigma \le q \le e+\sigma$ in
agreement with \cite{fisher89}.

\begin{figure}[h]
\centerline{\psfig{figure=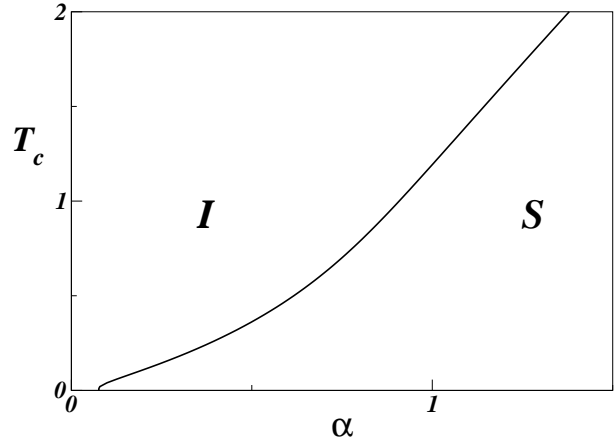,width=72mm,angle=270}}
\caption{Phase diagram for random offset charges with
probability distribution given by (\protect\ref{prob_delta_def}) and
short-ranged inverse capacitance matrix. In
the plot $p_0=p_e=1/2$.}
\label{figure4}
\end{figure}

We now consider the case 
\begin{equation}
P(q_{{\bf i}})=\sum_n p_n \delta(q_{{\bf i}}-ne),  
\label{prob_delta_def}
\end{equation}
with $\sum_n p_n=1$, corresponding to a random distribution of charges which
are integer multiples of $e$. Actually, this is the most realistic situation
for a random distribution. Indeed, the probability distributions employed
before should be viewed as fictitious continuous distributions; i.e., the
properties of the overall distribution of charges (mean value and width) can
be well approximated with a continuous distribution $P(q)$. Inserting the
probability distribution (\ref{prob_delta_def}) in (\ref{g_qy}) we have 
\[
\int dq P(q) g(q,y)= \int dq \sum_n p_n \delta(q-ne) g(q,y)=\sum_{odd} p_n
g(ne,y)+ \sum_{even} p_n g(ne,y) 
\]
where $\sum_{odd}$ ($\sum_{even}$) is a sum restricted to odd (even)
integer. Recalling Eq. (\ref{per}), we have $g(2ne,y)= g(0,y)$ and $%
g((2n+1)e,y)=g(e,y)$, and we find
\begin{equation}
\frac{1}{\alpha}=p_0 g(0,y)+p_e g(e,y)  \label{prob_delta}
\end{equation}
where $p_0=\sum_{even} p_n$ ($p_e=\sum_{odd} p_n$) is the probability that
the offset charge $q$ is an even (odd) integer multiple of $e$. In Fig.
\ref{figure4} we plot the phase boundary line (\ref{prob_delta}) for
$p_0=p_e=1/2$.

We are able to show now that applying the MF approximation described
in Appendix B with the probability distribution (\ref{prob_delta_def})
it is possible to find exactly Eq. (\ref{prob_delta}). The eigenvalue equation
for the Hamiltonian (\ref{QPM_mf}), 
$H_\ii \psi_n=E_n \psi_n$, can be recast in the standard form
of the Mathieu equation with the transformation $\psi_n(\varphi_{{\bf i}})=
e^{-i \frac{q_\ii}{2e} \varphi_{{\bf i}}} \rho_n(\varphi_{{\bf i}})$. We
find 
\[
\frac{d^2\rho_n}{d\varphi_{{\bf i}}^2}+\bigg(\frac{\lambda_n}{4}-\frac{v}{2}
\cos \varphi \bigg)\rho_n =0 
\]
where $\lambda_n=E_n /E_C$ and $v=- z E_J \overline{\langle\cos{\varphi}\rangle}/2 E_C$. 
The periodic boundary condition $\psi_n(\varphi_{{\bf i}})=
\psi_n(\varphi_{{\bf i}}+2\pi)$ gives $\rho_n(\varphi_{{\bf i}})=\rho_n(\varphi_{{\bf i}}+2
\pi)$ for $q_\ii=2ne$ and $\rho_n(\varphi_{{\bf i}})=-\rho_n(\varphi_{{\bf i}}
+2 \pi)$ for $q_\ii=2(n+1)e$ ($n$ integer). For small values of $v$, the
periodic and antiperiodic solutions of the Mathieu equation can be
calculated analytically, giving exactly Eq. (\ref{prob_delta}).

\subsection{NONDIAGONAL CAPACITANCE MATRICES}

\begin{figure}[h]
\centerline{\psfig{figure=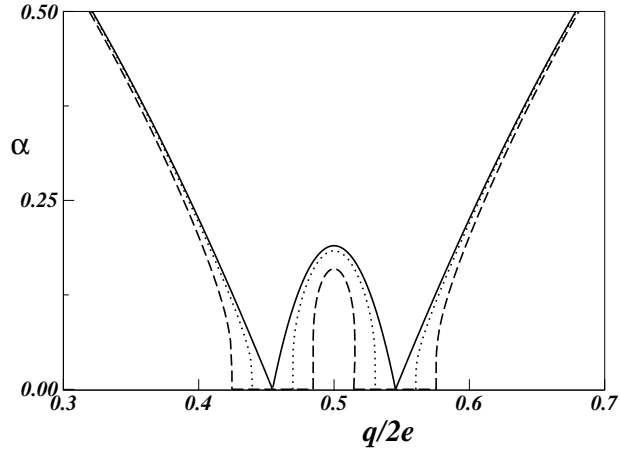,width=72mm,angle=270}}
\caption{Phase diagram at $T=0$ for random offset charges with 
uniform distribution and short-ranged inverse capacitance matrix. In
the plot $z\theta$ is equal to $0.1$ while $\sigma/2e$ is respectively
$0$ (solid line), $0.015$ (dotted line) and $0.03$ (dashed line). For this
value of $z\theta$, the lobe disappears at $\sigma=0.045$.}
\label{figure5}
\end{figure}

With nondiagonal capacitance matrices, the phase diagram without disorder
becomes richer \cite{fazio01}. For concreteness, we shall consider on-site
and a weaker nearest-neighbor (NN) interaction; i.e., the inverse
capacitance matrix is restricted to diagonal and NN terms. If one defines
$\theta$ as the ratio between NN and diagonal terms, one should restrict
only to $ z\theta<1$ in order to insure the invertibility of the
capacitance matrix \cite{fishman88}. Without disorder, at $T=0$ an
insulating lobe around $q=e$ appears: the width of this lobe is
$z\theta/(1+z\theta)$. Putting $ W=1+z\theta$, Eq. (\ref{g_qy}) for
$|q/2e|<1/2W$ gives $g(q,y \to 0)=1/[1-4W^2(q/2e)^2]$; for
$1/2W<q/2e<1-(1/2W)$ it becomes \cite{vanotterlo93}
\begin{equation}
g(q,y \to 0) = -\frac{1}{2}\left[\frac{1}{(2W\frac{q}{2e}-1) (2W\frac{q}{2e}
-3)}+ \frac{1}{(2W(\frac{q}{2e}-1)+1)(2W(\frac{q}{2e}-1)+3)}\right].
\end{equation}
In presence of disorder, Eq. (\ref{boundary-line-charge}) for a uniform
distribution gives $\alpha=0$ for $(1/2W)- \sigma \le q \le (1/2W)+ \sigma$
and $1- (1/2W)- \sigma\le q \le 1- (1/2W) + \sigma$. Thus, the lobe width
decreases as $(z\theta-2\sigma W)/W$. One sees that for $\sigma = z\theta/2W$
the insulating lobe around $q=e$ disappears. This phenomenon is evidenced in
Fig. \ref{figure5}.

\section{RANDOM SELF-CAPACITANCES}

In the following we limit our analysis only to JJA's with random
self-capacitance $C_{{\bf i} {\bf i}}$ and uniform charge frustration $q$.
We shall consider, in fact, a random diagonal charging energy 
$U_{{\bf i} {\bf i}}$, which we
assume to be independently distributed according to the probability
distribution $P(U_{{\bf i} {\bf i}}) \propto e^{-(U_{{\bf i} {\bf i}%
}-U_0)^2/2\sigma ^2}$, where the diagonal electrostatic contribution to the
energy $U_{{\bf i} {\bf i} }$ needs to be positive.
By averaging the free energy (\ref{av-ginzburg-landau}), the equation for
the phase boundary becomes
\begin{equation}  \label{boundary-line-energy}
\frac{1}{\alpha}=\int_0^{\infty} dU \frac{P(U)}{U} g(U,y)
\end{equation}
where now $\alpha=zE_J/4U_0$ and the function $g(U,y)$ is given by 
\begin{equation}  \label{g_qU}
g(U,y)=\frac{\sum_n e^{-\frac{4}{y}U(n+q/2e)^2} \frac{1}{1-4(n+q/2e)^2} } {%
\sum_{m}e^{-\frac{4}{y}U(m+q/2e)^2} }.
\end{equation}
The results of Eq. (\ref{boundary-line-energy}) are summarized in Figs. \ref%
{figure6} and \ref{figure7}: when $\sigma $ is small, the superconducting phase
increases in comparison to the nonrandom case: this is due to the factor $%
1/U$ in Eq. (\ref{boundary-line-energy}), which makes larger the
contribution of junctions with charging energies less than $U_{0}$. The
increase of the superconducting phase is thus due to a decrease of the
effective value of $E_{c}$. This behavior occurs until $\sigma $ reaches a
critical value (depending on the charge frustration and on the temperature),
of order $U_0$: at this value of $\sigma $ the insulating region starts to
increase. This is due to the asymmetry of the distribution, which has its
peak in $U_{0}$, but only for positive values. This phenomenon is present
also if one considers different distributions. An interesting observation is
that, when $q/2e=1/2$ (maximum frustration induced by the external offset
charges), the randomness does not modify considerably the phase diagram.
This should be compared with the nonfrustrated case ($q/2e=0$), where
randomness sensibly affects the phase diagram.

\section{CONCLUSIONS}

We obtained the phase diagram at finite temperature 
of JJA's with capacitive disorder (i.e., random offset charges and/or random
charging energies) by using a MF approximation. 
For a random distribution
of offset charges with mean $q$ and
variance $\sigma$, one has that for $\sigma \gtrsim e$, the phase boundary
line coincides for any value of $q$ and the lobe structures on the plane $
q-\alpha$ disappear ($\alpha$ is the ratio between the Josephson and
charging energies). At $T=0$ the result of Ref. \cite{fisher89} are
retrieved. If one considers in the BHM also a nearest-neighbor interaction,
the insulating lobe around $q=e$, which arises in absence of disorder, is
destroyed even for small values of $\sigma$. For arrays with random charging
energies, when the variance of the probability distribution is smaller than
a critical value, the superconducting phase increases with respect to the
situation in which all self-capacitances are equal. To check our results,
we have considered also the infinite-range hopping limit of the QPM.

Within the MF approach used here, 
it is not possible to capture the Bose glass phase 
and therefore the full physical picture 
can be much richer than the one extracted from MF.
However, if in low-dimensional systems (e.g., in
Josephson junction chains) the MF is expected to fail
\cite{bobbert92},
we envisage that our results provide qualitatively
correct predictions in large dimensional arrays.

{\bf Acknowledgements} We thank A.R. Bishop, F. Cooper, G. Grignani,
A. Mattoni, A. Smerzi, and A. Tagliacozzo for interesting and fruitful
discussions. A.T. and P.S. are grateful to the Los Alamos National
Laboratory for the kind hospitality and for partial financial support in the
final stage of this work. This research has been financed by M.I.U.R. under
grant No. 2001028294.

\appendix
\section{INFINITE-RANGE HOPPING LIMIT}

We consider the Hamiltonian (\ref{QPM}) in the limit of infinite-range
hopping, i.e., $\gamma _{{\bf i}{\bf j}}=1$ if ${\bf i}\neq {\bf j}$ and $%
\gamma _{{\bf i}{\bf j}}=0$ if ${\bf i}={\bf j}$. A meaningful thermodynamic
limit $(N\rightarrow \infty )$ is ensured by the scaling of the Josephson
term $E_{J}\equiv J/N$. This model has been studied in Ref. \cite{fisher89}
for the disordered BHM at $T=0$. Here we study the finite temperature case
within the approach provided in this paper. From $\gamma _{{\bf i}
{\bf j}}=\frac{1}{N}\sum_{{\bf k}}\gamma _{{\bf k}}e^{i{\bf k}\cdot ({\bf i}-
{\bf j})}$, it follows that
$\gamma _{{\bf k}}=N\delta _{{\bf k},0}-1$. 
Therefore the free energy (\ref{seconda_f}) is 
\begin{equation}
F[\psi ]=\frac{1}{\beta N}\sum_{\mu {\bf k}{\bf k}^{\prime }}\psi _{{\bf k}
}^{\ast }(\omega _{\mu })\left[ \frac{2N\delta _{{\bf k}{\bf k}^{\prime }}}{
J(N\delta _{{\bf k},0}-1)}-\frac{G_{{\bf k}-{\bf k}^{\prime }}(\omega _{\mu
})}{N}\right] \psi _{{\bf k}^{\prime }}(\omega _{\mu }).  \label{f_inf_range}
\end{equation}
Averaging over the disorder, requiring $\bar{F}=0$,  for $N\rightarrow
\infty $ one has

\begin{equation}
1=\frac{J}{2}\bar{G_{{\bf 0}}}.
\end{equation}

A comparison with Eq. (\ref{boundary-line}) makes evident that the results
of Secs. III and IV apply also in the infinite-range hopping model,
provided that $zE_{J}\rightarrow J$.

\section{AN ALTERNATIVE MEAN-FIELD APPROACH}

The effect of the quantum fluctuations is generally underestimated in
MF theory: 
we expect that a MF theory approach could 
provide qualitative information on the phase diagram in three dimensions,
but not in lower dimensions. 
In fact, a similar situation arises in quantum spin 
glasses where MF treatments are able to establish 
the existence of a spin glass transition \cite{bray80}.
Even if powerful and elegant, 
the mean-field approximation in the functional approach 
should be regarded only as a first step in understanding the 
role played by the disorder in the mean-field average which 
leads to Eq. (\ref{boundary-line}).

The aim of this appendix is to make more explicit
the relationship between the average over the disorder
and the quantum statistical average by resorting to a molecular
field theory approach. In fact,
we shall present an alternative and more 
intuitive mean-field approach, which may be easily applied
only to JJA's with diagonal capacitance matrices.

For this purpose, let us consider a JJA's with a fixed
charging energy $E_C$ and a random distribution of offset charges $P(q_\ii)$. 
The charging term in the QPM Hamiltonian (\ref{QPM}) is then diagonal and
the Josephson term couples different sites. Mean-field theory consists in
replacing the Josephson coupling of the phase on a given island $\ii$ 
to its neighbors by an average coupling: 
$E_{J}\sum_{\langle {\bf i} {\bf j}\rangle } 
\cos ({\varphi _{{\bf i}}}-{\varphi _{{\bf j}}})=
zE_{J}\overline{\langle \cos {\varphi }\rangle} 
\sum_{{\bf i}}\cos {\varphi _{{\bf i}}
}$. 
In this way the Hamiltonian becomes a sum of single site Hamiltonians, 
$H=\sum_{{\bf i}}H_{{\bf i}}$, where 
\begin{equation}
H_{{\bf i}}=-4E_{C}\frac{\partial ^{2}}{\partial \varphi _{{\bf i}}^{2}}%
-8iE_{C}\frac{q_{{\bf i}}}{2e}\frac{\partial }{\partial \varphi _{{\bf i}}}%
-zE_{J} \overline{\langle \cos {\varphi }\rangle} 
\cos {\varphi _{{\bf i}}}.
\label{QPM_mf}
\end{equation}
The single-site Hamiltonian (\ref{QPM_mf})
depends on the random charge frustration
$q_{{\bf i}}$: thus, its eigenfunctions and eigenvalues 
depend on $q_{{\bf i}}$. Therefore, one has to impose the self-consistency
condition with a double average, the quantum one and the average over the
disorder: 
\begin{equation}
\overline{\langle \cos \varphi \rangle }=\int dq_{{\bf i}}P(q_{{\bf i}})%
\frac{\sum_{n}e^{-\beta E_{n}}\langle \psi _{n}|\cos \varphi |\psi
_{n}\rangle }{\sum_{n}e^{-\beta E_{n}}}  \label{self-con}
\end{equation}
where the $\psi _{n}$ are the eigenfunctions of the single-site Hamiltonian
$H_i$ . The phase boundary line is
obtained from Eq. (\ref{self-con}) by  requiring
$\overline{\langle \cos \varphi \rangle}$ 
to be small and by keeping only terms proportional to it 
(we recall that the transition is second order).

We conclude this appendix by stressing that the application of mean-field
theory in the presence of disorder corresponds to introducing an order
parameter, which is averaged {\em also} over the disorder: the
self-consistency condition then gives the correct mean-field phase boundary
line. In Sec. III, Eqs. (\ref{self-con}) and (\ref{boundary-line})
are compared for the $\delta$-like probability
distribution, showing that they give the same results.

\begin{figure}
\centerline{\psfig{figure=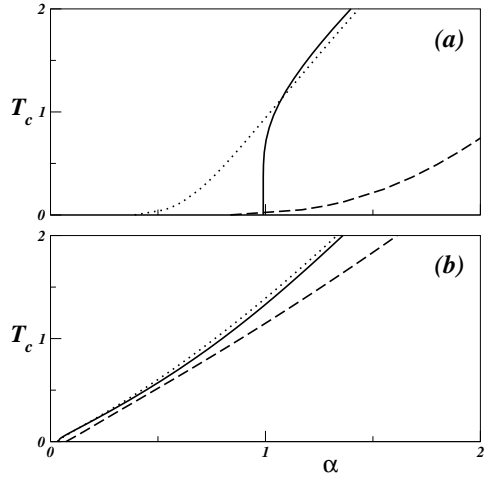,width=72mm,angle=270}}
\caption{Phase diagram in the $T_c-\alpha$ plane for random diagonal
capacitance with Gaussian distribution and uniform offset charge
$q/2e=0$ (a), $0.5$ (b) ($T_c$ is in units of
$k_B/U_0$) while $\sigma/2e$ is $0.1$ (solid line), $1$ (dotted line) and
$5$ (dashed line).} \label{figure6}
\end{figure}

 \begin{figure}
\centerline{\psfig{figure=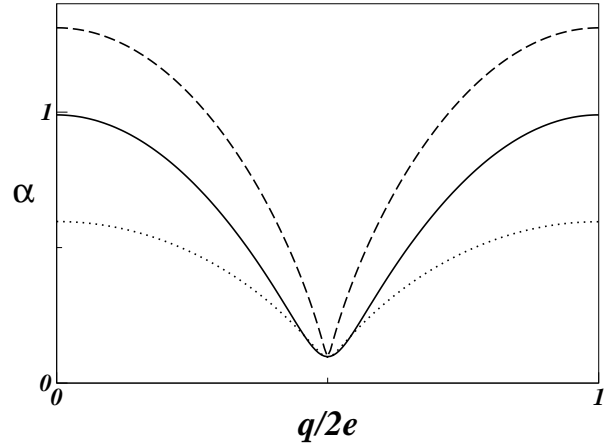,width=72mm,angle=270}}
\caption{Phase diagram in the $q-\alpha$   plane for random diagonal
capacitance with Gaussian distribution and uniform offset charge
at $k_B T /U_0=0.1$.
$\sigma/2e=0.1$ (solid line), $1$ (dotted line) and $5$ (dashed line).}
\label{figure7}
\end{figure}
\end{document}